\documentclass[12pt,preprint]{aastex}
\shorttitle{INTERNAL ABSORPTION OF GAMMA RAYS IN 3C 279}

\shortauthors{Liu et al.}

\begin{document}

\title{ABSORPTION OF 10 GeV--1 TeV GAMMA RAYS BY RADIATION FROM BROAD-LINE REGION IN 3C 279}

\author{H. T. Liu\altaffilmark{1}, J. M. Bai\altaffilmark{1}, and L. Ma\altaffilmark{2}}

\altaffiltext{1}{National Astronomical Observatories/Yunnan
Astronomical Observatory, Chinese Academy of Sciences, Kunming,
Yunnan 650011, China; liuhongtao1111@hotmail.com;
baijinming@ynao.ac.cn}

\altaffiltext{2} {Physics Department, Yunnan Normal University,
Kunming 650092, China; send offprint requests to
liuhongtao1111@hotmail.com}

\begin{abstract}

In this paper, we study the photon-photon pair production optical
depth for 10 GeV--1 TeV gamma rays from 3C 279 due to the diffuse
radiation of broad-line region (BLR). Assuming a power-law
spectrum of $E_{\gamma}^{-a_2}$ for the photon intensity of very
high energy (VHE) gamma rays, $a_1 \gtrsim 405$ and $a_2\gtrsim
6.4$ are inferred by the integrated photon fluxes measured by
MAGIC and HESS. Based on this power-law spectrum, the pre-absorbed
spectra are inferred by correcting the photon-photon absorption on
the diffuse photons of the BLR (internal absorption) and the
extragalactic background light (external absorption). Position of
gamma-ray emitting region $R_{\rm{\gamma}}$ determines the
relative contributions of this two diffuse radiation to the total
absorption for 10 GeV--1 TeV gamma rays. The internal absorption
could make spectral shape of gamma rays more complex than only
corrected for the external absorption, and could lead to the
formation of arbitrary softening and hardening gamma-ray spectra.
It should be necessary for the internal absorption to be
considered in studying 10 GeV--1 TeV gamma rays from powerful
blazars. The energies of annihilated gamma-ray photons due to the
internal absorption are likely to be mainly reradiated around GeV.
Our results indicate that $R_{\rm{\gamma}}$ may be between the
inner and outer radii of the BLR for 3C 279. This implies for
powerful blazars that $R_{\rm{\gamma}}$ might be neither inside
the BLR cavity nor outside the BLR, but be within the BLR shell.
Observations of $\it GLAST$, MAGIC, HESS, and VERITAS in the near
future could give more constraints on the position of the
gamma-ray emitting region relative to the BLR.

\end{abstract}

\keywords{gamma rays: theory --- quasars: individual (3C 279)}

\section{INTRODUCTION}
The classical flat spectrum radio quasar (FSRQ) 3C 279 is one of
the brightest extragalactic objects in the gamma-ray sky. It was
detected by the EGRET, and its spectrum does not show any
signature of gamma-ray absorption by pair production up to $\sim$
10 GeV (Fichtel et al. 1994; von Montigny et al. 1995). With the
detection of high energy gamma rays in 66 blazars, containing 51
FSRQs and 15 BL Lac objects, in the GeV energy range by the EGRET
experiment aboard the Compton Gamma Ray Observatory (Catanese et
al. 1997; Fichtel et al. 1994; Lin et al. 1997; Mukherjee et al.
1997; Thompson et al. 1995, 1996; Villata et al. 1997; Hartman et
al. 1999; Nolan et al. 2003), an exceptional opportunity is
presented for the understanding of the central engine operating in
blazars. Some of blazars have been also firmly detected by
atmospheric Cerenkov telescopes at energies above 1 TeV, such as
Mrk 421 (Punch et al. 1992), and Mrk 501 (Quinn et al. 1996). The
High Energy Stereoscopic System (HESS) is an imaging atmospheric
Cerenkov detector with the energy threshold above 100 GeV (Funk et
al. 2004; Hinton 2004; Hofmann 2003). The Very Energetic Radiation
Imaging Telescope Array System (VERITAS) provides unprecedented
sensitivity to photon energies between 50 GeV and 50 TeV (see
e.g., Holder et al. 2006). The Major Atmospheric Gamma Imaging
Cerenkov telescope (MAGIC) is currently the largest single-dish
Imaging Air Cerenkov Telescope in operation with the lowest energy
threshold, $\sim$ 30 GeV, among the new Cerenkov telescopes (see
e.g., Baixeras et al. 2004). At present, 23 active galactic nuclei
(AGNs) have been detected in very high energy (VHE) gamma rays,
containing 21 BL Lac objects, one radio source M87, and the first
FSRQ 3C 279 with the highest redshift $z=0.536$ in these VHE
AGNs\footnote{http://www.mppmu.mpg.de/$\sim$rwagner/sources/}. The
MAGIC telescope detected VHE gamma rays from 3C 279 (Teshima et
al. 2007). HESS observations measured an upper limit of integrated
photon flux (Aharonian et al. 2008). The Large Area Telescope
instrument on the Gamma-Ray Large Area Space Telescope ($\it
GLAST$), the new-generation high energy gamma-ray telescope with
sufficient angular resolution to allow identification of a large
fraction of their optical counterparts, will observe gamma rays
with energies from 20 MeV to greater than 300 GeV, and have the
unique capability to detect thousands of gamma-ray blazars to
redshifts of at least $z=4$ (see e.g., Chen et al. 2004). $\it
GLAST$ was launched on June 11 2008. $\it GLAST$, combined with
new-generation TeV instruments such as MAGIC, HESS, and VERITAS,
will tremendously improve blazar spectra studies, filling in the
band from 20 MeV to 10 TeV with high significance data for
hundreds of AGNs (see Gehrels \& Michelson 1999). Future
measurements of the gamma-ray spectrum shape and its variability
of blazars would tremendously improve our understanding to
blazars.

These gamma rays from blazars are generally believed to be
attributed by emission from a relativistic jet oriented at a small
angle to the line of sight (Blandford \& Rees 1978). These
gamma-ray components are contributed by inverse Compton emissions,
including synchrotron self-Compton (SSC) scattering synchrotron
seed photons, and external Compton (EC) scattering seed photons
from sources outside the jet (see e.g., B\"ottcher 1999). The
diffuse radiation fields of broad-line region (BLR) could have a
strong impact on the expected EC spectra of powerful blazars,
FSRQs (Liu \& Bai 2006; Reimer 2007; Tavecchio \& Ghisellini
2008). The external soft photon fields not only provide target
photons for the EC processes to produce these gamma-ray
components, but also absorb gamma rays from the EC processes,
because gamma rays between 10 GeV and 1 TeV interact with
infrared-ultraviolet photons to be attenuated by photon-photon
pair production. Many efforts to study the absorption of gamma
rays focus on the photon-photon annihilation by the diffuse
extragalactic background radiation at the infrared (IR), optical,
and ultraviolet (UV) bands (see e.g., Stecker et al. 1992, 2006,
2007; Stecker \& de Jager 1998; Oh 2001; Renault et al. 2001; Chen
et al. 2004; Dwek \& Krennrich 2005; Schroedter 2005). This
external absorption of gamma rays on the diffuse extragalactic
background light (EBL) is also proposed and used to probe the EBL
(Renault et al. 2001; Chen et al. 2004; Dwek \& Krennrich 2005;
Schroedter 2005). Indeed, the internal absorption of gamma rays
inside FSRQs could result in serious problem for the possibility
to use the external absorption of gamma rays to probe the
IR--optical--UV extragalactic background, because the intrinsic
spectra of gamma rays are masked by the internal absorption inside
blazars (Donea \& Protheroe 2003; Liu \& Bai 2006; Reimer 2007).
The intrinsic spectra of gamma rays are complicated by the complex
spectra of the diffuse radiation fields of the BLRs in FSRQs
(Tavecchio \& Ghisellini 2008). The mean of intrinsic spectral
index is around $2.3$ for 17 BL Lac objects detected in the VHE
regime (Wagner 2008).

The positions of gamma-ray emitting regions are still an open and
controversial issue in the researches on blazars. It is suggested
that gamma rays are produced inside the BLRs and the gamma-ray
emitting radius $R_{\gamma}$ ranges roughly between 0.03 and 0.3
pc (Ghisellini \& Madau 1996). It is argued by Georganopoulos et
al. (2001) that the radiative plasma in relativistic jets of
powerful blazars are within cavities formed by the BLRs. However,
other researchers argued that the gamma-ray emitting regions are
outside the BLRs (Lindfors et al. 2005; Sokolov \& Marscher 2005).
In our previous research (Liu \& Bai 2006, hereafter Paper I), the
position of gamma-ray emitting region is a key parameter to
determine whether high energy gamma rays could escape the diffuse
radiation fields of the BLRs for FSRQs. It is unknown whether
these gamma rays could be detected by $\it GLAST$, MAGIC, HESS,
and VERITAS even if blazars intrinsically produce 10 GeV--1 TeV
gamma rays, because the gamma-ray emitting radii are unknown.
These gamma rays around 200 GeV is optically thick for 3C 279 if
the emitting region is within the BLR cavity (see Paper I).
However, the MAGIC observations on 2006 February 23 have shown a
clear gamma-ray signal in the VHE regime (Teshima et al. 2007).
This indicates that the emitting region of these VHE gamma rays
should not be inside the BLR cavity, otherwise, the intrinsic flux
of these VHE gamma rays is likely to be extremely high. In Paper
I, we addressed an important topic in gamma-ray astrophysics,
namely the absorption of high energy gamma rays inside of FSRQs by
photons of the BLR. In this paper, we attempt to address the
particular topic of absorption in the gamma-ray quasar 3C 279
using the available observational data, and its potential effect
on the spectra of gamma rays. In order to constrain the position
of gamma-ray emitting region in 3C 279, we study the internal
absorption of gamma rays by the diffuse radiation from the BLR and
its potential effect on the spectra of gamma rays from 10 GeV to 1
TeV.

The structure of this paper is as follows. $\S$ 2 presents
intensity of VHE gamma rays. $\S$ 3 presents theoretical
calculation, and consists of two subsections. $\S$ 3.1 presents
calculations of temperature profiles, and $\S$ 3.2 photon-photon
optical depth for 3C 279. $\S$ 4 presents external absorption on
IR--optical--UV extragalactic background. $\S$ 5 is spectral shape
of VHE gamma rays. $\S$ 6 presents pair spectrum due to
photon-photon annihilation and their radiation. $\S$ 7 is for
discussions and conclusions. Throughout this paper, we use a flat
cosmology with a deceleration factor $q_0=0.5$ and a Hubble
constant $H_0=75 \/\ \rm{km \/\ s^{-1} \/\ Mpc^{-1}}$.

\section{INTENSITY OF VHE GAMMA RAYS}
The classical FSRQ 3C 279 is one of the brightest extragalactic
objects in the gamma-ray sky, and it is also the first VHE
gamma-ray FSRQ. For 3C 279, there are ten MAGIC observation nights
from 2006 January 31 to 2006 March 31. While in most of the nights
gamma-ray fluxes compatible with zero were observed, a marginal
signal was seen during the 2006 February 22 observations. In the
night of 2006 February 23, observations have shown a clear
gamma-ray signal with an integrated photon flux
$F(E_{\gamma}>200\/\ \rm{GeV})=(3.5\pm0.8)\times 10^{-11}\/\
\rm{cm^{-2}s^{-1}}$ (Teshima et al. 2007). This gamma-ray
detection was not accompanied by an optical flare or by
particularly high flux levels or outbursts in X-rays (Teshima et
al. 2007). After one day, 3C 279 brightens from a marginal signal
to a clear signal detected by MAGIC. This suggests that the VHE
gamma-ray emissions may have variations on the order of days. The
GeV gamma-ray emissions detected by the EGRET have variations on
timescale as short as 1 day (Hartman et al. 2001a). This variation
timescale is roughly consistent with that observed by MAGIC. This
agreement indicates that the GeV and VHE gamma rays may have some
certain relationship. However, no VHE spectrum of 3C 279 has been
published yet.

For 3C 279, observations on 2006 February 22--23 show the optical
R magnitude around 14.5 (see Fig. 1 in B\"ottcher et al. 2007),
and a flux density around 5--5.2 $\rm{mJy}$ in the R band (see
Fig. 1 in Teshima et al. 2007). The average of R band magnitude on
2000 February 8 to March 1 is around 14.5 (see Fig. 2 in
Kartaltepe \& Balonek 2007). The simultaneous multi-wavelength
observations on 2000 February 8 to March 1 for 3C 279 show a flux
density from 4.2 to 8.9 $\rm{mJy}$ in the R band (Hartman et al.
2001b). These R band magnitudes and flux densities observed on
2000 February 8 to March 1 are consistent with those observed on
2006 February 22--23. This agreement indicates that the gamma-ray
emissions on this two periods are likely to have very similar
states. Thus, the gamma-ray emissions on 2006 February 22--23 are
likely to be at mediate states. The gamma-ray emissions in most of
the nights when gamma-ray fluxes compatible with zero were
observed are likely to be at low states.

Though, an integrated photon flux is given by Teshima et al.
(2007), no VHE spectrum of 3C 279 has been published. Fortunately,
HESS observations in 2007 January measured an upper limit of
integrated photon flux $F(E_{\gamma}>300\/\ \rm{GeV})<3.98\times
10^{-12}\/\ \rm{cm^{-2}s^{-1}}$ (Aharonian et al. 2008).
Observations show that VHE spectra of these known VHE sources can
be described by a power-law spectrum (see Wagner 2008). Thus, a
power-law spectrum of $\frac{dI}{dE_{\gamma}}=a_1
E_{\gamma}^{-a_2}$ is assumed for the photon intensity of the VHE
gamma rays. This VHE gamma-ray spectrum could be limited by the
two measured integrated photon fluxes. Taking into account the
integrated photon fluxes measured by MAGIC and HESS: adopting
$F(E_{\gamma}>200\/\ \rm{GeV})=3.5\times 10^{-11}\/\
\rm{cm^{-2}s^{-1}}$ and assuming $F(E_{\gamma}>300\/\
\rm{GeV})=3.9\times 10^{-12}\/\ \rm{cm^{-2}s^{-1}}$, the photon
intensity of the VHE gamma rays is inferred as
\begin{equation}
\frac{dI}{dE_{\gamma}}=505 \times E_{\gamma}^{-6.4}\/\
\rm{cm^{-2}s^{-1}GeV^{-1}}.
\end{equation}
Though, $a_2\gtrsim 6.4$ is limited by the HESS observations, we
will adopt $a_2=6.4$ in the following calculations, and will
discuss the potential effect of $a_2\gtrsim 6.4$ on our results in
the last section. The spectral slope of $6.4$ is larger than that
of any other known VHE sources (see Wagner 2008). The maximum
spectral slope of $4.21$ has been measured in the VHE regime for
PG 1553+113 (Albert et al. 2007). This spectral slope of $6.4$ is
also larger than the slope of $2.3$ for the extrapolated VHE
spectrum by the GeV spectrum observed on 2000 February 8 to March
1 (Hartman et al. 2001b). The average EGRET blazar spectrum was
found to have a slope of $2.27$ (Venters \& Pavlidou 2007). The
mean of intrinsic spectral indices is $\overline{\Gamma}\approx
2.3$ for 17 blazars detected in the VHE regime (see Wagner 2008).
The reason that this spectral slope of 6.4 is larger than any
other known photon index of gamma rays, measured and intrinsic
spectral indices, is likely to be the internal and external
absorption of gamma rays by the diffuse radiation fields of the
BLR and the EBL. Stecker et al. (1992) investigated the
photon-photon absorption of the VHE gamma-ray spectrum,
extrapolated by the differential spectrum of gamma rays measured
by EGRET during 1991 June, on the extragalactic background
infrared radiation field. The corrected photon flux is not
inconsistent with the Whipple upper limit (Stecker et al. 1992).
This indicates that the VHE and GeV gamma rays are likely to have
some relationship. In the following section, we shall firstly
calculate the internal absorption within 3C 279 by the diffuse
radiation fields produced by the BLR.

\section{THEORETICAL CALCULATION}
As described in $\S$ 2.1, 2.2, and 2.3 of Paper I, equations
(1)--(21) can be used to estimate the absorption optical depth by
the diffuse radiation of the BLR, adopting the spherical shell of
clouds (see Fig. 1 in Paper I) and the relative intensity of broad
emission lines presented on Figure 2 in Paper I, and assuming a
blackbody temperature in the inner regions of accretion disk. The
assumption of the single-temperature blackbody probably differs
significantly from the real cases. The surface effective
temperatures of accretion disks are functions of radii
$r_{\rm{d}}$, i.e., $T_{\rm{eff}}=T_{\rm{eff}}(r_{\rm{d}})$ (see
e.g., Ebisawa et al. 1991; Hanawa 1989; Li et al. 2005; Pereyra et
al. 2006; Shakura \& Sunyaev 1973; Zimmerman et al. 2005). These
continua from thin accretion disks can be well described by
multi-temperature blackbody. Considering the temperature profile
of accretion disk, the factor $n_{\rm{bb}}(\nu,T)$ in equations
(20) and (21) in Paper I should be replaced by
\begin{equation}
n_{\rm{bb}}(\nu,T)=\int^{X_{\rm{out}}}_{X_{\rm{in}}}n_{\rm{bb}}(\nu,T_{\rm{eff}}(X_{d}))
dX_{\rm{d}},
\end{equation}
where $X_{\rm{in}}=r_{\rm{in}}/r_{\rm{g}}$ and
$X_{\rm{out}}=r_{\rm{out}}/r_{\rm{g}}$ are the inner and outer
radii of accretion disk, respectively, and
$X_{\rm{d}}=r_{\rm{d}}/r_{\rm{g}}$. The gravitational radius of a
black hole with mass of $M_{\rm{BH}}$ is $r_{\rm{g}}=G
M_{\rm{BH}}/c^2$, where $G$ is the gravitational constant and $c$
is the speed of light.

In the calculations of the photon-photon attenuation optical
depth, the soft photon frequency is considered in the range from
$\nu^L_2=10^{12.0}\/\ \rm{Hz}$ to $\nu^U_2=10^{16.5}\/\ \rm{Hz}$
for the diffuse continuum from the BLR. The temperature profile of
accretion disk is considered for standard thin accretion disk.

\subsection{Calculations of Temperature Profiles}
The local effective temperatures of accretion disks are functions
of radii $r_{\rm{d}}$ (see e.g., Ebisawa et al. 1991; Hanawa 1989;
Li et al. 2005; Pereyra et al. 2006; Shakura \& Sunyaev 1973;
Zimmerman et al. 2005). The standard accretion disk is the basic
model for a radiatively efficient, geometrically thin disk.
Variability of active galactic nuclei is mostly in favor of the
standard accretion disk models of AGNs (Liu et al. 2008).
Although, equations required to calculate the temperature profile
are the same as equations (5)--(12) in our previous paper (Liu et
al. 2008), they are presented as equations (A1)--(A8) to make this
paper more readable (see Appendix A). For high luminosity blazars
showing a clear UV bump, the ironing luminosity
$L_{\rm{iron}}=L_{\rm{UV}}=\eta \dot{M}c^2$ according to the
definition of the efficiency $\eta$ with which various types of
black holes convert rest mass-energy into outgoing radiation
(Thorne 1974). From the observed BLR luminosity $L_{\rm{BLR}}$ and
relation $L_{\rm{BLR}}=f_{\rm{cov}}L_{\rm{UV}}$ (D'Elia et al.
2003), we could estimate the mass accretion rate of the central
black hole by the formula
\begin{equation}
\dot{M}=\frac{L_{\rm{UV}}}{\eta_{\rm{max}}c^2}=\frac{L_{\rm{BLR}}}{\eta_{\rm{max}}f_{\rm{cov}}c^2}.
\end{equation}
The temperature profiles could be estimated by equations
(A1)--(A8) and equation (3).

The dimensionless spin parameter of a black hole can take on any
value in the range $-1\leq a_{\rm{\ast}} \leq 1$, where negative
values of $a_{\rm{\ast}}$ correspond to a black hole that
retrogrades relative to its accretion disk. For simplicity we
consider only prograde spins up to the Thorne spin equilibrium
limit, i.e. $0\leq a_{\rm{\ast}}\leq 0.998$ (Thorne 1974). Recent
work on magnetohydrodynamic accretion disks suggest a rather lower
equilibrium spin (see e.g., Gammie et al. 2004; Krolik et al.
2005). Spin equilibrium is reached at $a_{\rm{\ast}}\approx 0.93$
through accretion of gases onto the central black hole, and
mergers of black holes with comparable mass can result in a final
spin of $a_{\rm{\ast}}\sim $ 0.8--0.9 (Gammie et al. 2004).
Equilibrium spins as low as $a_{\rm{\ast}}\sim 0.9$ are within the
realm of possibility (Krolik et al. 2005). Aschenbach et al.
(2004) obtained a value of
$a_{\rm{\ast}}=0.9939^{+0.0026}_{-0.0074}$ for the Galactic Center
black hole. Brenneman \& Reynolds (2006) obtained a formal
constraint on spin $a_{\rm{\ast}}=0.989^{+0.009}_{-0.002}$ at 90\%
confidence for the Seyfert galaxy MCG--06-30-15. Considering the
probable ranges of spin parameter $a_{\rm{\ast}}$ suggested above,
we take three values of spin parameter $a_{\rm{\ast}}=0.5$, $0.8$,
and $0.998$ in the Kerr metric to calculate the temperature
profiles. Combining equations (A1)--(A8), equation (3), and the
parameters of $M_{\rm{BH}}$, $L_{\rm{BLR}}$, $f_{\rm{cov}}$, and
$a_{\rm{\ast}}$, the surface effective temperature profiles are
calculated. The calculated results are presented in Figure 1.

\subsection{Photon-Photon Optical Depth For 3C 279}
As stated in $\S 5$ of Paper I, the absorption of gamma rays by
emission lines is unrelated to the BLR covering factor
$f_{\rm{cov}}$, and the gamma-ray absorption by the diffuse
blackbody radiation is in proportion to the ratio
$\tau_{\rm{BLR}}/f_{\rm{cov}}$, where $\tau_{\rm{BLR}}$ is the
Thomson optical depth of BLR. Various values of the BLR covering
factor are suggested and estimated. Early estimates of this
quantity indicated a covering factor $f_{\rm{cov}}\sim 5-10\%$,
while recent observations indicated $f_{\rm{cov}}\sim 30\%$ (see
e.g., Maiolino et al. 2001). D'Elia et al. (2003) found only two
blazars in literatures: 3C 273 and PKS 2149-306, of which both UV
continuum emission from accretion disks and broad emission lines
have been measured. They obtained $f_{\rm{cov}}\sim 7\%$ for the
first source and $f_{\rm{cov}}\sim 10\%$ for the second one. Donea
\& Protheroe (2003) suggested $f_{\rm{cov}}\sim 3\%$ for the
spherical distribution of the BLR clouds in quasars. The issue of
the Thomson optical depth of BLR has rarely been studied and has
gotten nowhere for blazars. Blandford \& Levison (1995) adopted
the Thomson optical depth of BLR as $\tau_{\rm{BLR}}=0.01$. Thus,
the ratio of $\tau_{\rm{BLR}}/f_{\rm{cov}}$ is likely to be
typically of order $\tau_{\rm{BLR}}/f_{\rm{cov}}\sim 1$ for the
spherical distribution of the BLR clouds in blazars. In the
calculations of $\tau_{\gamma\gamma}$, we adopted
$\tau_{\rm{BLR}}/f_{\rm{cov}}=1$ and $f_{\rm{cov}}=0.03$ for the
spherical distribution of the BLR clouds in FSRQs.

3C 279 has a BLR luminosity $L_{\rm{BLR}}=10^{44.41}\/\ \rm{ergs
\/\ s^{-1}}$ (Cao \& Jiang 1999), and a black hole mass of
$M_{\rm{BH}}=10^{8.4}\/\ M_{\sun}$ (Woo \& Urry 2002). It was
detected by the EGRET in the 0.1--10 GeV energy domain, and its
spectrum can be fitted by a power law without any signature of
gamma-ray absorption by pair production (Fichtel et al. 1994; von
Montigny et al. 1995). Firstly, we investigated the dependence of
the effective temperature $T_{\rm{eff}}$ on the central black hole
spin $a_{\rm{\ast}}$ for 3C 279. The inner radii of accretion
disks in the Kerr potential are fixed at the marginally stable
orbit
$r_{\rm{ms}}(a_{\rm{\ast}})=X_{\rm{ms}}(a_{\rm{\ast}})r_{\rm{g}}$.
The outer radii of accretion disks are fixed at $r_{\rm{out}}=200
r_{\rm{g}}$. The temperature profiles of the standard accretion
disks are presented in Figure 1. It is obvious that the
temperatures for three different values of $a_{\rm{\ast}}$ are
nearly the same at the radius $r_{\rm{out}}=200 r_{\rm{g}}$. The
variations of $a_{\rm{\ast}}$ can significantly change the
temperature profiles (see Fig. 1). It is evident that the
different temperature profiles produce different multi-temperature
blackbody continua, and then could result in different absorption
for gamma rays.

We calculated $\tau_{\gamma\gamma}$ for 3C 279 by adopting the BLR
inner radius $r_{\rm{BLR,in}}=0.1 \/\ \rm{pc}$ and the BLR outer
radius $r_{\rm{BLR,out}}=0.4 \/\ \rm{pc}$ from Hartman et al.
(2001b). It is suggested that the gamma-ray emitting region should
be within the BLR cavity (Ghisellini \& Madau 1996; Georganopoulos
et al. 2001). Firstly, we assume $R_{\rm{\gamma}}=r_{\rm{BLR,in}}$
to calculate the absorption optical depth. The absorption optical
depth is presented in Figure 2. It can be seen in Figure 2 that
the diffuse radiation fields of the BLR are not transparent to
gamma rays of energies from 10 GeV to 1 TeV in the observer frame
if these gamma rays are inside the BLR cavity. This result is
inconsistent with observations that show no signature of gamma-ray
absorption by pair production around 10 GeV (see e.g., Fichtel et
al. 1994; von Montigny et al. 1995; Ghisellini \& Madau 1996;
Wehrle et al. 1998). Thus for $\sim$ 10 GeV gamma rays,
observations should not support suggestions of gamma-ray emitting
region within the BLR cavity. It is likely that the gamma-ray
emitting region is within the BLR, i.e.
$r_{\rm{BLR,in}}<R_{\rm{\gamma}}<r_{\rm{BLR,out}}$. The absorption
optical depth is calculated by assuming
$R_{\rm{\gamma}}=(r_{\rm{BLR,out}}+r_{\rm{BLR,in}})/2$ (see Fig.
3). It is obvious that the VHE gamma rays are still significantly
absorbed by the diffuse radiation of the BLR (only $\sim$ 5\%
escape probability). These 10 GeV gamma rays are still slightly
absorbed. The slightly larger cover factor $f_{\rm{cov}}$ could
make it vanish for this slight absorption around 10 GeV. Thus, the
gamma-ray emitting region is likely to be within the BLR. The
gamma-ray emitting regions outside the BLRs are argued by some
researchers (Lindfors et al. 2005; Sokolov \& Marscher 2005).
Therefore, $R_{\rm{\gamma}}=r_{\rm{BLR,out}}$ is assumed to
calculate the absorption optical depth. The calculated results are
presented in Figure 4. There is no absorption for 10 GeV gamma
rays (see Fig. 4). This is consistent with the EGRET observations.
For the VHE gamma rays, greater than $\sim$ 60\% of the primary
gamma rays can escape the diffuse radiation fields of the BLR (see
Fig. 4).

\section{EXTERNAL ABSORPTION ON IR-OPTICAL-UV EXTRAGALACTIC BACKGROUND}
Many efforts are focused on investigating the EBL at
IR--optical--UV bands, and the photon-photon annihilation
absorption of gamma rays by the EBL at these bands (see e.g.,
Stecker et al. 1992; Stecker \& de Jager 1998; Oh 2001; Renault et
al. 2001; Chen et al. 2004; Dwek \& Krennrich 2005; Schroedter
2005). The external absorption of gamma rays by the diffuse EBL is
also proposed and used to probe the EBL (Renault et al. 2001; Chen
et al. 2004; Dwek \& Krennrich 2005; Schroedter 2005). Stecker et
al. (1992) first pointed out the importance of the EBL in
determining the opacity of the universe to high energy gamma rays
at higher redshifts. Dwek \& Krennrich (2005) detailed the
observational limits and detections of the EBL, and the relevant
EBL spectral templates. Their investigations showed that the
absorption of gamma rays $\lesssim$ 1 TeV is entirely contributed
by the EBL at 0.1--10 $\rm{\mu m}$ (see Fig. 3 in Dwek \&
Krennrich 2005). Dwek \& Krennrich only calculated the optical
depth for low redshift sources. Stecker et al. (2006, 2007) have
given the optical depth for sources at redshifts $<6$. Stecker et
al. (1992) investigated the IR EBL absorption on high energy gamma
rays for 3C 279, and got absorption optical depth of $3.7 \lesssim
\tau \lesssim 9.7$ for 1 TeV gamma rays. An analytic form to
approximate the function $\tau_{\gamma\gamma}(E_{\gamma},z)$ of
the external EBL absorption is given as (Stecker et al. 2006)
\begin{equation}
\log \tau_{\gamma\gamma}=Ax^4+Bx^3+Cx^2+Dx+E,
\end{equation}
where $x\equiv\log E_{\gamma}(\rm{eV})$. Coefficients A through E
are given in Table 1 for various redshifts (Stecker et al. 2006,
2007). The parametric approximation holds for $10^{-2}\leq
\tau_{\gamma\gamma} \leq 10^2$ and $E_{\gamma}\lesssim 2 \/\
\rm{TeV}$. These coefficients at redshift $z=0.5$ for the baseline
model fit are adopted to calculate the EBL absorption optical
depth.

In order to compare the internal and external absorption, the
calculated results for the external absorption are also presented
in Figures 2$a$, 3$a$, and 4$a$. If $R_{\rm{\gamma}}$ is around
the inner radius $r_{\rm{BLR,in}}$, the internal absorption
dominates over the external absorption, and the latter mainly
presents itself in the VHE interval and is much less than unity
around 10 to a few ten GeV (see Fig. 2$a$). The internal
absorption peaks around 200 GeV, and the external absorption
increases with energies of gamma rays (see Fig. 2$a$). If
$R_{\rm{\gamma}}$ is around the median of $r_{\rm{BLR,in}}$ and
$r_{\rm{BLR,out}}$, the total absorption is dominated by the
internal absorption in the interval of 10--100 GeV, and the
external absorption dominates over the internal one from 300 GeV
to 1 TeV (see Fig. 3$a$). The relative contributions of the
internal and external absorption to the total one are comparable
around 200 GeV (see Fig. 3$a$). If $R_{\rm{\gamma}}$ is around the
outer radius $r_{\rm{BLR,out}}$, the internal absorption is
comparable with the external one from 10 to 20 GeV, and the former
is dominated by the latter in the interval of 30 GeV to 1 TeV (see
Fig. 4$a$). The positions of gamma-ray emitting regions determine
the relative contributions of the internal and external absorption
to the total photon-photon annihilation optical depth for 10
GeV--1 TeV gamma rays.

\section{SPECTRAL SHAPE OF VHE GAMMA RAYS}
Measured and intrinsic VHE gamma-ray speactra can be well
described by a power-law spectrum, and intrinsic gamma-ray spectra
of 17 BL Lac objects are inferred by only correcting the EBL
absorption to the measured spectra (e.g., Wagner 2008). If the
gamma-ray emitting regions are far from the BLRs in FSRQs, it is
reasonable to infer the intrinsic spectra by only correcting the
EBL absorption. Otherwise, it is likely to be insufficient for
FSRQs to infer the intrinsic spectra from the measured VHE spectra
by only correcting the EBL absorption, because the internal
absorption is not negligible when compared with the external
absorption (see Figs. 2$a$ and 3$a$). The positions of gamma-ray
emitting regions determine the relative contributions of the
internal and external absorption. The dependence of the internal
absorption on energies of gamma rays relies on the positions of
gamma-ray emitting regions (see Figs. 2$a$, 3$a$, and 4$a$). The
external absorption monotonically increases with energies of gamma
rays, and then its effect on the spectral shape is more
straightforward than that of the internal absorption. It is
obvious that the external absorption softens the observed
gamma-ray spectrum relative to the emission one.

In order to study the dependence of the internal absorption on the
gamma-ray emitting radius $R_{\gamma}$, we accepted gamma-ray
energy $E_{\gamma}=$ 50, 100, 300, 500, and 1000 GeV to calculate
the photon-photon absorption optical depth. The calculated results
are presented in Figure 5. It is obvious that the absorption
optical depth $\tau_{\gamma\gamma}$ rapidly decreases with
increasing $R_{\gamma}$, because the energy density of the diffuse
radiation of the BLR rapidly decreases beyond $r_{\rm{BLR,in}}$.
For a fixed $E_{\gamma}$, $\tau_{\gamma\gamma}$ monotonically
decreases as $R_{\gamma}$ increases. For a fixed $R_{\gamma}$, the
dependence of $\tau_{\gamma\gamma}$ on $E_{\gamma}$ relies on
$R_{\gamma}$. The internal absorption optical depth
$\tau_{\gamma\gamma}$ of VHE gamma rays does not monotonically
vary with $E_{\gamma}$, and peaks around a few hundred GeV if
$R_{\gamma}\lesssim 0.35 \/\ \rm{pc}$ (see Fig. 5). This is
confirmed by these calculated results presented in Figures 2$a$
and 3$a$. The optical depth $\tau_{\gamma\gamma}$ for any
$E_{\gamma}$ monotonically varies with $E_{\gamma}$ if
$R_{\gamma}\gtrsim 0.35 \/\ \rm{pc}$ (see Figs. 4$a$ and 5). The
optical depth $\tau_{\gamma\gamma}$ of high energy gamma rays
monotonically increases with $E_{\gamma}$ for any $R_{\gamma}$
(see Figs. 2$a$, 3$a$, 4$a$, and 5).

After gamma-ray spectra are corrected for the internal and
external absorption, the pre-absorbed spectra are given for three
values of $R_{\gamma}$ (see dash-dot-dotted curves in Figs. 2$b$,
3$b$, and 4$b$). If $R_{\gamma}$ is around $r_{\rm{BLR,in}}$, the
pre-absorbed VHE gamma-ray spectra peak around 200 GeV (see Fig.
2$b$). The internal absorption hardens the VHE gamma-ray spectrum
around 300 GeV--1 TeV, and softens the gamma-ray spectrum below
200 GeV (see Fig. 2$b$). If $R_{\gamma}$ is around the median of
inner and outer radii of the BLR, the internal absorption softens
the gamma-ray spectrum below $\sim$ 400 GeV, and hardens the VHE
gamma-ray spectrum from $\sim$ 400 GeV to 1 TeV (see Fig. 3$b$).
If $R_{\gamma}$ is beyond the outer radius of the BLR, the
internal absorption softens the gamma-ray spectrum (see Fig.
4$b$), and has the same effect as the external absorption. The
external absorption softens these gamma-ray spectra that escape
from the diffuse radiation fields of the BLR. After passing
through the internal and external diffuse radiation fields, these
detected gamma-ray spectra are softer than those pre-absorbed ones
below $\sim$ 400 GeV, and are harder than those pre-absorbed ones
from $\sim$ 400 GeV to 1 TeV, when $R_{\gamma}$ is around
$R_{\rm{BLR,in}}$ (see Fig. 2$b$). As $R_{\gamma}$ is beyond the
median of inner and outer radii of the BLR, these detected
gamma-ray spectra are softer than those pre-absorbed ones (see
Figs. 3$b$ and 4$b$). If the intrinsic spectral indices have a
typical value of 2.3 for VHE gamma-ray sources, the first VHE FSRQ
3C 279 may have an intrinsic spectral index around 2.3 in the VHE
regime. These pre-absorbed VHE gamma rays from 200 GeV to 1 TeV
can be well fitted by a power-law spectrum with a chance
probability of $p<10^{-20}$. As $R_{\gamma}=0.1 \/\ \rm{pc}$,
photon indices of $5.7\pm0.2$ ($a_{\ast}=0.5$) and $4.9\pm0.2$
($a_{\ast}=0.998$) are given by fit. These values are larger than
the typical value of $2.3$ for those known VHE sources, and are
also larger than the intrinsic photon index $\simeq 3.6$ for PG
1553+113, the maximum among those known VHE spectra (Wagner 2008).
Photon indices of $1.97\pm0.009$ ($a_{\ast}=0.5$) and
$1.94\pm0.003$ ($a_{\ast}=0.998$) are given for $R_{\gamma}=0.25
\/\ \rm{pc}$. These two values are smaller than the typical value
of $2.3$ in the VHE regime, but are larger than the intrinsic
photon index $\simeq 1.3$ for 1ES 1101-232, the minimum among
those known VHE spectra (Wagner 2008). Photon indices of
$1.7\pm0.03$ are given for $R_{\gamma}=0.4 \/\ \rm{pc}$. This
value is smaller than the typical value of $2.3$, but is larger
than the minimum of 1.3. If $a_2$ is allowed to rise, these fit
photon indices for the pre-absorbed VHE spectra could be
increased. Thus, it is unlikely for 3C 279 that these gamma-ray
emitting regions are inside the BLR cavity.

\section{PAIR SPECTRUM DUE TO PHOTON-PHOTON ANNIHILATION AND THEIR RADIATION}
In this section, our efforts are focused on studying where
energies of annihilated gamma-ray photons are likely to be
reradiated. B\"ottcher \& Schlickeiser (1997) derived the exact
analytic solution of pair production spectrum from photon-photon
annihilation, and showed that this exact solution is in very good
agreement with the approximation of Aharonian et al. (1983), the
most accurate one of the various approximations known before. The
interaction of power-low gamma-ray spectra with thermal soft
photon fields is generally described within an error of a few
percent at all electron/position energies if the soft photon
temperature is $kT/m_{\rm{e}}c^2 \lesssim 0.1$, even if gamma-ray
spectra extend down to $\varepsilon_1 \sim 1$. Thus, we adopted
the approximation of Aharonian et al. (see eq. [32] of B\"ottcher
\& Schlickeiser 1997) to estimate the pair injection rate
$\frac{dn(\gamma)}{dt}$ due to photon-photon absorption by the
diffuse radiation fields of the BLR in 3C 279. A power-law
spectrum is assumed as $\propto E_{\gamma}^{-2.3}$ for the
pre-absorbed gamma-ray spectrum in order to calculate the pair
injection rate. For simplicity in the calculations of
$\frac{dn(\gamma)}{dt}$, we adopted radial independent soft photon
densities with such a relative intensity that can produce the
comparable quantity of $\int \frac{dn(\gamma)}{dt} d\gamma$,
converted from the energies of annihilated gamma-ray photons, for
the diffuse multi-temperature blackbody and broad emission lines.
After the soft photon densities and the pair production rate
$\frac{dn(\gamma)}{dt}$ are given, the equilibrium pair
distribution of steady state could be estimated. Figure $6a$ shows
the differential pair injection rate $\frac{dn(\gamma)}{dt}$ for
the interaction of a power-law gamma-ray spectrum $\propto
E_{\gamma}^{-2.3}$ with the BLR diffuse radiation of 3C 279. The
$\gamma$ values of electrons/positrons annihilated from the broad
emission lines and gamma rays have a lower limit of $\gamma \ga
10^4$ (see Fig. 6$a$). The $\gamma$ values of electrons/positrons
from annihilation between the multi-temperature blackbody and
gamma rays have a lower limit of $\gamma \ga 10^3$ (see Fig.
6$a$).

If considering the radiative cooling, synchrotron cooling and
external Compton scattering cooling, the equilibrium pair
distribution of steady state is
$n(\gamma)=\int^{\gamma_{\rm{max}}}_{\gamma}
\frac{dn(\gamma)}{dt}d \gamma /\dot{\gamma}$, where
$\dot{\gamma}=\dot{\gamma}_{\rm{syn}}(\gamma)+\dot{\gamma}_{\rm{EC}}(\gamma)$,
and $\gamma =E_{\rm{e^{\mp}}}/m_{\rm{e}}c^2$ is electron/positron
energy. The synchrotron cooling rate is given by the formula
$\dot{\gamma}_{\rm{syn}}(\gamma)=\frac{4}{3}\frac{\sigma_{\rm{Th}}}{m_e
c}\frac{B^2}{8\pi}\gamma^2$, where $B$ is the magnetic field
intensity. The EC scattering cooling rate is
$\dot{\gamma}_{\rm{EC}}(\gamma)=\int d\varepsilon_1 \varepsilon_1
\int F_{\rm{KN}}(\varepsilon_1,\varepsilon_2,\gamma)
n(\varepsilon_2)d\varepsilon_2$, where we use the Compton kernel
$F_{\rm{KN}}$ for an isotropic soft photons, considering the full
Klein-Nishina cross section (Jones 1968; Blumenthal \& Gould
1970). The Compton kernel is
\begin{equation}
F_{\rm{KN}}=\frac{3}{4}\frac{c\sigma_{\rm{Th}}}{\varepsilon_2
\gamma^2}\left[2q\ln
q+1+q-2q^2+\frac{(\gamma_{\varepsilon_2}q)^2(1-q)}{2(1+\gamma_{\varepsilon_2}q)}\right],
\end{equation}
where $\gamma_{\rm{\varepsilon_2}}=4\varepsilon_2\gamma$,
$q=\frac{\varepsilon_1}{4\varepsilon_2
\gamma(\gamma-\varepsilon_1)}$, and $1/4\gamma^2<q<1$ (Jones 1968;
Blumenthal \& Gould 1970). $\varepsilon_1$ and $\varepsilon_2$ in
this paper are energies of gamma rays and soft photons in units of
$m_ec^2$, respectively. The synchrotron emission coefficient is
\begin{equation}
j_{\rm{syn}}(\nu)=\frac{\sqrt{3}e^3B}{4\pi m_e c^2}\int d\gamma
n(\gamma)R_{\rm{CS}}(x),
\end{equation}
where the mean emission coefficient for a single electron/positron
averaged over an isotropic distribution of pitch angles
$R_{\rm{CS}}(x)$ (Crusius \& Schlickeiser 1986; Ghisellini et al.
1988) is
\begin{equation}
R_{\rm{CS}}(x)=2x^2
K_{4/3}(x)K_{1/3}(x)-\frac{6x^3}{5}\left[K_{4/3}^2(x)-K_{1/3}^2(x)\right],
\end{equation}
where $x=\nu/3\gamma^2\nu_{\rm{B}}$ and $\nu_{\rm{B}}=eB/2\pi
m_{\rm{e}} c$, and $K_{\rm{n}}$ is the McDonald function of order
$n$ (see also Saug\'e \& Henri 2004). The EC emission coefficient
$j_{\rm{EC}}(\nu)$ is
\begin{equation}
j_{\rm{EC}}(\nu_1)=\frac{h}{4\pi}\varepsilon_1\int \int
F_{\rm{KN}}(\varepsilon_1,\varepsilon_2,\gamma)n(\varepsilon_2)n(\gamma)d\gamma
d\varepsilon_2,
\end{equation}
where $n(\varepsilon_2)$ is the differential soft photon density.
For 3C 279, we adopted the magnetic field intensity $B=1.5 \/\
\rm{G}$ from Hartman et al. (2001b). The calculated spectra are
presented in Figure 6$b$. For FSRQs, the ratio of the EC
luminosity to synchrotron luminosity is
$L_{\rm{EC}}/L_{\rm{syn}}\sim 10$, on average (Ghisellini et al.
1998; Georganopoulos et al. 2001). The relative radiation
intensities of the diffuse blackbody and broad emission lines of
the BLR and the magnetic field intensity, adopted in calculations,
could produce the ratio of $L_{\rm{EC}}/L_{\rm{syn}}= \nu
J_{\rm{\nu,EC}}/ \nu J_{\rm{\nu,syn}} \sim 10$ for the pair
spectrum due to photon-photon absorption (see Fig. 6$b$). It can
be seen in Figure 6$b$ that the synchrotron emissions peak around
keV, and the EC emissions peak around GeV. Thus, the intense
creation of pairs would produce a strong radiation at low energy
X-rays and around GeV energy. These produced gamma rays around
GeV, where the pileup of photons below the absorption threshold
occurs, could result in a significant flattening in the observed
spectrum relative to the emission spectrum.

\section{DISCUSSIONS AND CONCLUSIONS}
HESS observations in 2007 January measured an upper limit of
integrated photon flux $F(E_{\gamma}>300\/\ \rm{GeV})<3.98\times
10^{-12}\/\ \rm{cm^{-2}s^{-1}}$ (Aharonian et al. 2008). In
calculations, we assumed $F(E_{\gamma}>300\/\ \rm{GeV})=3.9\times
10^{-12}\/\ \rm{cm^{-2}s^{-1}}$, allowed by HESS observations.
Combining this flux with integrated photon flux
$F(E_{\gamma}>200\/\ \rm{GeV})=3.5\times 10^{-11}\/\
\rm{cm^{-2}s^{-1}}$ measured by MAGIC (Teshima et al. 2007), a
power-law spectrum of $\frac{dI}{dE_{\gamma}}=a_1
E_{\gamma}^{-a_2}$ is inferred with $a_1 =505$ and $a_2 =6.4$. If
$F(E_{\gamma}>300\/\ \rm{GeV})=3.6\times 10^{-12}\/\
\rm{cm^{-2}s^{-1}}$ is adopted for HESS observations, $a_1 =1510$
and $a_2 =6.6$. If $F(E_{\gamma}>300\/\ \rm{GeV})=3.98\times
10^{-12}\/\ \rm{cm^{-2}s^{-1}}$ is adopted, $a_1 =405$ and $a_2
=6.36$. Thus $a_1 \gtrsim 405$ and $a_2 \gtrsim 6.4$ are likely to
be limited by MAGIC and HESS observations for a VHE gamma-ray
spectrum of $\frac{dI}{dE_{\gamma}}=a_1 E_{\gamma}^{-a_2}$. Based
on the gamma-ray spectrum of $\frac{dI}{dE_{\gamma}}=505
E_{\gamma}^{-6.4}$ but corrected for the internal and external
absorption (see Figs. 2$b$, 3$b$, and 4$b$), photon indices of
these corrected VHE gamma rays are around 5 as $R_{\gamma}=0.1 \/\
\rm{pc}$, 2.0 as $R_{\gamma}=0.25 \/\ \rm{pc}$, and 1.7 as
$R_{\gamma}=0.4 \/\ \rm{pc}$. These inferred photon indices as
$R_{\gamma}=$ 0.25 and 0.4 $\rm{pc}$ are inside of those known
intrinsic photon indices. If $a_2 =6.6$, photon indices of
$1.9\pm0.03$ are inferred for $R_{\gamma}=0.4 \/\ \rm{pc}$,
$2.2\pm0.01$ ($a_{\ast}=0.5$) and $2.1\pm 0.003$
($a_{\ast}=0.998$) for $R_{\gamma}=0.25 \/\ \rm{pc}$, and
$6.0\pm0.2$ ($a_{\ast}=0.5$) and $5.1\pm0.2$ ($a_{\ast}=0.998$)
for $R_{\gamma}=0.1 \/\ \rm{pc}$. As $R_{\gamma}$ varies between
0.1 and 0.4 $\rm{pc}$, photon indices of pre-absorbed VHE
gamma-ray spectra are likely to be inside of intrinsic photon
index range from 1.3 to 3.6 (Wagner 2008). If one smaller
$F(E_{\gamma}>300\/\ \rm{GeV})=2.0\times 10^{-12}\/\
\rm{cm^{-2}s^{-1}}$ is adopted, one larger $a_2=8.0$ is obtained.
As $a_2 =8.0$, photon indices of $3.3\pm0.03$ are inferred for
$R_{\gamma}=0.4 \/\ \rm{pc}$, $3.6\pm0.01$ ($a_{\ast}=0.5$) and
$3.5\pm 0.003$ ($a_{\ast}=0.998$) for $R_{\gamma}=0.25 \/\
\rm{pc}$, and $7.3\pm0.2$ ($a_{\ast}=0.5$) and $6.5\pm0.2$
($a_{\ast}=0.998$) for $R_{\gamma}=0.1 \/\ \rm{pc}$. Thus, larger
photon indices of these pre-absorbed spectra are obtained for a
fixed $R_{\gamma}$ as larger $a_2$ is adopted. As
$R_{\gamma}=r_{\rm{BLR,in}}$, photon indices of these pre-absorbed
gamma-ray spectra are always larger than the typical value
$\Gamma_{\rm{in}}=2.3$ and the maximum $\Gamma_{\rm{in}}=3.6$ of
those known intrinsic photon indices. Thus, it is unlikely for 3C
279 that $R_{\gamma}$ be inside the BLR cavity, i.e. it is likely
$R_{\gamma}> r_{\rm{BLR,in}}$. If $F(E_{\gamma}>300\/\ \rm{GeV})$
adopted is closer to the upper limit of HESS observations, $a_2$
is not too large. Too large $a_2$, such as $a_2\simeq 10$
corresponding to $F(E_{\gamma}>300\/\ \rm{GeV})\simeq 10^{-12}\/\
\rm{cm^{-2}s^{-1}}$, seems to be impossible for VHE gamma-ray
spectra because these spectra with $a_2 \gtrsim 10$ are much
steeper (softer) than the steepest (softest) spectrum measured for
PG 1553+113 (Albert et al. 2007). When gamma-ray emitting region
is already beyond the BLR, the EC mechanisms, where external
photons originate from the BLR, are made inefficient to produce
the observed gamma rays. For 3C 279, soft photon energy density
around $r_{\rm{BLR,out}}$ is lower by a factor of 6 to 7 than that
around $r_{\rm{BLR,in}}$ for the BLR diffuse radiation produced by
the spherical shell of clouds (see Fig. 1 in Paper I). Thus it is
unlikely $R_{\gamma}>r_{\rm{BLR,out}}$, i.e. it is likely
$R_{\gamma}\lesssim r_{\rm{BLR,out}}$.

The external absorption softens the observed spectra relative to
the emission ones in the interval from 10 GeV to 1 TeV. Whether
the internal absorption softens or hardens the observed spectra
relies on gamma-ray emitting radius $R_{\gamma}$ and energy
$E_{\gamma}$. Photon index variations of gamma-ray spectra
relative to $a_2=6.4$ are calculated for the internal absorption,
and the internal and external absorption (see Figure 7). As
$R_{\gamma}=r_{\rm{BLR,in}}$, the local photon indices of
gamma-ray spectra decrease with $E_{\gamma}$ from 10 to 110 GeV,
and increase with $E_{\gamma}$ from 110 GeV to 1 TeV (see Figure
7$a$). The local photon indices of gamma-ray spectra corrected for
the internal absorption recover around 270 GeV ($\Delta
\Gamma(\rm{in})=0$). The internal absorption make gamma-ray
spectra softer and softer from 10 to 110 GeV and from 270 down to
110 GeV, and make gamma-ray spectra harder and harder from 270 GeV
to 1 TeV. After corrected for the internal and external
absorption, the local photon indices of gamma rays recover around
500 GeV ($\Delta \Gamma(\rm{in+ext})=0$). The external absorption
can not change the trend of the local photon indices of gamma-ray
spectra corrected for the internal absorption. As
$R_{\gamma}=(r_{\rm{BLR,in}}+r_{\rm{BLR,out}})/2$, the local
photon indices of gamma-ray spectra corrected for the internal
absorption behave as in the case of $R_{\gamma}=r_{\rm{BLR,in}}$
(see Figure 7$b$). These local photon indices recover around 400
GeV ($\Delta \Gamma(\rm{in})=0$). The internal absorption make
gamma-ray spectra softer and softer from 10 to 110 GeV and from
400 down to 110 GeV, and make gamma-ray spectra harder and harder
from 400 GeV to 1 TeV. After corrected for the internal and
external absorption, the local photon indices of pre-absorbed
gamma-ray spectra basically decrease with $E_{\gamma}$. The
external absorption can not change the trend of the local photon
indices of gamma-ray spectra corrected for the internal absorption
below 100 GeV, but change that above 100 GeV. Gamma-ray spectra
become softer and softer relative to pre-absorbed ones from 10 GeV
to 1 TeV ($\Delta \Gamma(\rm{in+ext})<0$). As
$R_{\gamma}=r_{\rm{BLR,out}}$, the local photon indices have
similar trend to that in the case of
$R_{\gamma}=(r_{\rm{BLR,in}}+r_{\rm{BLR,out}})/2$ (see Figure
7$c$). The internal absorption make gamma-ray spectra softer and
softer from 10 to $\sim$ 420 GeV and from 1 TeV down to $\sim$ 420
GeV. After corrected for the internal and external absorption, the
local photon indices monotonically decrease with $E_{\gamma}$, and
gamma-ray spectra become softer and softer relative to
pre-absorbed ones from 10 GeV to 1 TeV ($\Delta
\Gamma(\rm{in+ext})<0$). The external absorption change the trend
of the local photon indices of gamma-ray spectra corrected for the
internal absorption. The calculated results presented in Figure 7
basically are independent on the particular value of $a_2$. For
example, the trends of the local photon indices for $a_2=5.4$ are
basically identical to those in the case of $a_2=6.4$. In summary,
the internal absorption could make spectral shape more complex
than only considering the external absorption, and could lead to
the formation of arbitrary softening and hardening gamma-ray
spectra (see Figs. 2--4 and 7). Thus, it should be necessary for
the internal absorption to be considered in studying gamma rays of
10 GeV--1 TeV from FSRQs.

Assuming $\Gamma_{\rm{jet}}=15$ for 3C 279, most of gamma rays are
contained within a radiation cone with a half open angle of
$\Delta \varphi\sim 1/ \Gamma_{\rm{jet}} \sim 3.8^{\circ}$,
because of the relativistic beaming effect. If the central
IR--optical--UV photons coming directly from accretion disks
travel through the radiation cone, the IR--optical--UV photons can
have photon-photon pair creation processes with gamma rays within
the radiation cone. The gamma-ray emitting region
$R_{\rm{\gamma}}\sim 0.1 \/\ \rm{pc}$ is assumed, and the radii of
the UV radiation regions are $R_{D}\lesssim 30 r_{\rm{g}}\sim
0.0004 \/\ \rm{pc}$ (see Fig. 1), the angle between the jet
direction and the travelling direction of UV photons at
$R_{\rm{\gamma}}$ is $\sim \arcsin
R_{\rm{D}}/R_{\rm{\gamma}}\lesssim 0.2^{\circ}$. Then the photons
within the radiation cone have the collision angles of $\theta
\lesssim 4.0^{\circ}$. For the UV photons at the frequency $\nu
\simeq10^{16.5} \/\ \rm{Hz}$ with energies of $\varepsilon_2\simeq
2.56\times 10^{-4}$ and the gamma rays of $\varepsilon_1\simeq
2.0\times 10^{6}$ corresponding to energies around 1 TeV, the left
of threshold condition (eq. [3] in Paper I) has a upper limit of
$\lesssim 0.6$, which is less than unity, and thus the two kinds
of photons cannot be absorbed by the photon-photon pair creation
processes. Therefore, the central UV radiation have negligible
contributions to the absorption for gamma rays relative to the
diffuse radiation from the BLR. For optical photons, the radii of
the optical radiation regions are $R_{D}< 200 r_{\rm{g}}\sim 0.003
\/\ \rm{pc}$, the angle between the jet direction and the
travelling direction of optical photons at $R_{\rm{\gamma}}$ is
$\sim \arcsin R_{\rm{D}}/R_{\rm{\gamma}}\lesssim 1.7^{\circ}$.
Then the photons within the radiation cone have the collision
angles of $\theta \lesssim 5.5^{\circ}$. For the optical photons
at the frequency $\nu \simeq10^{15} \/\ \rm{Hz}$ with energies of
$\varepsilon_2\simeq 8.1\times 10^{-6}$ and the gamma rays of
$\varepsilon_1\simeq 2.0\times 10^{6}$, the left of threshold
condition has a upper limit of $\lesssim 0.04$, which is less than
unity, and thus the two kinds of photons cannot be absorbed by the
photon-photon pair creation processes. Therefore, the central
optical radiation have negligible contributions to the absorption
for gamma rays relative to the diffuse radiation from the BLR. For
IR photons, the central radiation also have negligible
contributions to the absorption for gamma rays. Thus, the central
IR--optical--UV radiation have negligible contributions to the
absorption for gamma rays relative to the diffuse radiation from
the BLR.

Absorption for gamma rays by photon-photon annihilation and where
the energies carried by the annihilated gamma rays reradiate are
important to gamma-ray research. The intense creation of pairs
would produce a strong radiation at low energy X-rays or at GeV
energy (Protheroe \& Stanev 1993; Saug\'e \& Henri 2004; Zdziarski
\& Coppi 1991). The electron-positron pair cascade could cause the
soft X-ray excesses (Zdziarski \& Coppi 1991). The produced
electron-positron pairs could make a difference around 1 GeV,
where the pileup of photons below the absorption threshold results
in a significant flattening in the observed spectrum relative to
the emission spectrum (Protheroe \& Stanev 1993). In $\S$ 6, we
studied the pair spectrum due to photon-photon absorption, and the
synchrotron and EC spectra emitted by the equilibrium pair
distribution of steady state. The synchrotron radiation peaks
around keV X-rays, and the EC radiation peaks around GeV gamma
rays (see Fig. 6$b$). Thus, pairs due to annihilation absorption
for gamma rays by the diffuse radiation fields of the BLR are
likely to make a difference around 1 GeV for 3C 279.

In this paper, in order to limit the gamma-ray emitting radius
$R_{\gamma}$, we used a BLR model to study the photon-photon
absorption by the diffuse radiation of the BLR in 3C 279 for gamma
rays of 10 GeV to 1 TeV in the observed spectrum. We calculated
the internal absorption of gamma rays from 10 GeV to 1 TeV for
$R_{\gamma}=r_{\rm{BLR,in}}$, $r_{\rm{BLR,out}}$, and
$(r_{\rm{BLR,in}}+r_{\rm{BLR,out}})/2$ (see Figs. 2$a$, 3$a$, and
4$a$). For a fixed $R_{\gamma}$, dependence of photon-photon
absorption optical depth $\tau_{\gamma\gamma}$ on energies of
gamma rays $E_{\gamma}$ relies on $R_{\gamma}$. Dependence of
$\tau_{\gamma\gamma}$ on $R_{\gamma}$ was also studied for a fixed
$E_{\gamma}$ (see Fig. 5). For a fixed $E_{\gamma}$,
$\tau_{\gamma\gamma}$ decreases with increasing $R_{\gamma}$. The
external absorption on the IR--optical--UV EBL was also estimated
for gamma rays of 10 GeV--1 TeV, and it monotonically increases as
$E_{\gamma}$ increases. Comparing the internal absorption with the
external one shows that $R_{\gamma}$ determines the relative
contributions of the internal and external absorption to the total
photon-photon annihilation absorption of 10 GeV--1 TeV gamma rays
(see Figs. 2$a$, 3$a$, and 4$a$). Based on MAGIC and HESS
observations, a power-law spectrum as in equation (1) was adopted
for the photon intensity of VHE gamma rays. The pre-absorbed
gamma-ray spectra are inferred by this power-law corrected for the
internal and external absorption. The internal absorption could
make spectral shape of gamma rays more complex than that only
corrected for the external absorption, and could lead to the
formation of arbitrary softening and hardening gamma-ray spectra
(see Figs. 2$a$, 3$a$, 4$a$, and 7). Thus, it should be necessary
for the internal absorption to be considered in studying 10 GeV--1
TeV gamma rays from FSRQs. $R_{\gamma}$ significantly influences
the variations of spectral shape due to the internal absorption.
Calculations imply that the energies of annihilated gamma rays due
to the internal absorption are mainly reradiated around GeV regime
(see Fig. 6$b$). Considering the possible variations of photon
index $a_2$, the photon indices of the pre-absorbed VHE gamma-ray
spectra were compared with those known intrinsic photon indices.
As $R_{\gamma}=r_{\rm{BLR,in}}$ and $a_2\gtrsim 6.4$, the photon
indices of the pre-absorbed gamma-ray spectra are always larger
than the typical value $\Gamma_{\rm{in}}=2.3$ and the maximum
$\Gamma_{\rm{in}}=3.6$ of those known intrinsic photon indices,
and $\tau_{\gamma\gamma}$ is larger than unity. Thus, it is likely
$R_{\gamma}> r_{\rm{BLR,in}}$ for 3C 279. As
$R_{\gamma}=r_{\rm{BLR,out}}$ and $6.4 \lesssim  a_2 \lesssim
8.3$, photon indices of pre-absorbed gamma-ray spectra are not
larger than $\Gamma_{\rm{in}}=3.6$. For a fixed $a_2$, photon
indices of pre-absorbed gamma-ray spectra decrease as $R_{\gamma}$
increases. Too large $a_2$ seems to be impossible for VHE
gamma-ray spectra. In addition, the EC processes may be
inefficient to produce the observed gamma rays as gamma-ray
emitting region is already beyond the BLR. Thus, it is likely
$R_{\gamma}\lesssim r_{\rm{BLR,out}}$ for 3C 279. Our results
suggest that $R_{\gamma}$ for powerful blazars might be neither
inside the BLR cavity nor outside the BLR, but be within the BLR
shell. This is neither consistent with suggestions of Ghisellini
\& Madau (1996) and Georganopoulos et al. (2001) nor consistent
with suggestions of Lindfors et al. (2005) and Sokolov \& Marscher
(2005).

Our results are model dependent, especially dependent on the
assumed power-law spectrum for the VHE gamma rays. If Teshima et
al. published the spectral indices of gamma-ray spectra measured
by MAGIC (Teshima et al. 2007), the power-law spectrum assumed in
this paper could be tested. Tavecchio \& Ghisellini (2008) used
the photoionization code CLOUDY, described by Ferland et al.
(1998), to calculate the detailed spectra from the BLRs for
powerful blazars, and then used these spectra to calculate the EC
spectra. Approximate spectra of the BLRs are used in this paper
and Paper I, and another approximate spectra are used by Reimer
(2007). Difference between the detailed and approximate spectra
should exist. It should be useful in the future researches to
study the effects of this difference on the results of previous
researches using approximate spectra (e.g., Liu \& Bai 2006;
Reimer 2007). Observations of $\it GLAST$, MAGIC, HESS, and
VERITAS in the near future could give more observational
constraints on the gamma-ray emitting regions and the BLRs for the
powerful blazars. Publications of intrinsic photon indices
predicted by theoretical researches and photon indices measured by
observations in the VHE regime could give stronger constraints on
$R_{\gamma}$.

\acknowledgements We are grateful to the anonymous referee for
his/her constructive comments and suggestions leading to
significant improvement of this paper. H. T. L. thanks for
financial support by National Natural Science Foundation of China
(NSFC, Grant 10573030) and (Grant 10778726). L. M. is supported by
NSFC (Grant 10778702). J. M. B. thanks support of the Bai Ren Ji
Hua project of the CAS, China.
\appendix

\section{Appendix}
If the central black holes are Kerr ones, the local effective
temperature of the standard disk is given in the Kerr metric as
(Krolik 1999)
\begin{equation}
T_{\rm{eff}}(X_{\rm{d}})=\left[ \frac{3G M_{\rm{BH}} \dot{M}}{8
\pi \sigma_{\rm{SB}} r_{\rm{g}}^3 X_{\rm{d}}^3}
R_{\rm{R}}(X_{\rm{d}})\right] ^{1/4},
\end{equation}
where $\sigma_{\rm{SB}}$ is the Stefan-Boltzmann constant,
$M_{\rm{BH}}$ is the black hole mass, $\dot{M}$ is the mass
accretion rate of the central black hole, and the function
$R_{\rm{R}}(X_{\rm{d}})$ is
\begin{equation}
R_{\rm{R}}(X_{\rm{d}})=\frac{C(X_{\rm{d}})}{B(X_{\rm{d}})},
\end{equation}
where the functions $B(X_{\rm{d}})$ and $C(X_{\rm{d}})$ are,
respectively, (Krolik 1999)
\begin{equation}
B(X_{\rm{d}})=1-\frac{3}{X_{\rm{d}}}+\frac{2a_{\rm{\ast}}}{X_{\rm{d}}^{3/2}},
\end{equation}
\begin{eqnarray}
C(X_{\rm{d}}) &=&
1-\frac{y_{\rm{ms}}}{y}-\frac{3a_{\rm{\ast}}}{2y}\ln\left(\frac{y}{y_{\rm{ms}}}\right)
 \nonumber\\
 &&-\frac{3(y_1-a_{\rm{\ast}})^2}{yy_1(y_1-y_2)(y_1-y_3)}\ln\left(\frac{y-y_1}{y_{\rm{ms}}-y_1}\right)
 \nonumber\\
 && -\frac{3(y_2-a_{\rm{\ast}})^2}{yy_2(y_2-y_1)(y_2-y_3)}\ln\left(\frac{y-y_2}{y_{\rm{ms}}-y_2}\right)
 \nonumber\\
 && -\frac{3(y_3-a_{\rm{\ast}})^2}{yy_3(y_3-y_1)(y_3-y_2)}\ln\left(\frac{y-y_3}{y_{\rm{ms}}-y_3}\right) \;,
\end{eqnarray}
where $y={(X_{\rm{d}})}^{1/2}$, $a_{\rm{\ast}}=cJ/GM^2_{\rm{BH}}$
is the dimensionless spin parameter of the central black hole with
the spin angular momentum $J$, $y_{\rm{ms}}={(X_{\rm{ms}})}^{1/2}$
is the value of $y$ at the marginally stable orbit, and
$y_{1,2,3}$ are the three roots of $y^3-3y+2a_{\rm{\ast}}=0$ (see
e.g., Reynolds \& Nowak 2003).

Assuming prograde orbits, the radii of the marginally stable
orbits in the equatorial plane of a Kerr black hole are (Bardeen
et al. 1972)
\begin{equation}
X_{\rm{ms}}=3+Z_2-\left[(3-Z_1)(3+Z_1+2Z_2) \right]^{1/2},
\end{equation}
where
\begin{equation}
Z_1=1+\left(1-a^2_{\rm{\ast}}\right)^{1/3}\left[\left(1+a_{\rm{\ast}}\right)^{1/3}+
\left(1-a_{\rm{\ast}}\right)^{1/3}\right],
\end{equation}
and
\begin{equation}
Z_2=\left(3a^2_{\rm{\ast}}+Z_1^2\right)^{1/2}.
\end{equation}
The marginally stable orbits in the equatorial plane correspond to
the maximum efficiency of energy release as a result of accretion,
assuming prograde orbits (Kembhavi \& Narlika 1999)
\begin{equation}
\eta_{\rm{max}}=1-\frac{X_{\rm{ms}}-2+a_{\rm{\ast}}X_{\rm{ms}}^{-1/2}}{\sqrt{X_{\rm{ms}}\left(
X_{\rm{ms}}-3+2a_{\rm{\ast}}X_{\rm{ms}}^{-1/2}\right)}}.
\end{equation}

\clearpage

\begin{figure}
\centering
\includegraphics[angle=-90,scale=0.5]{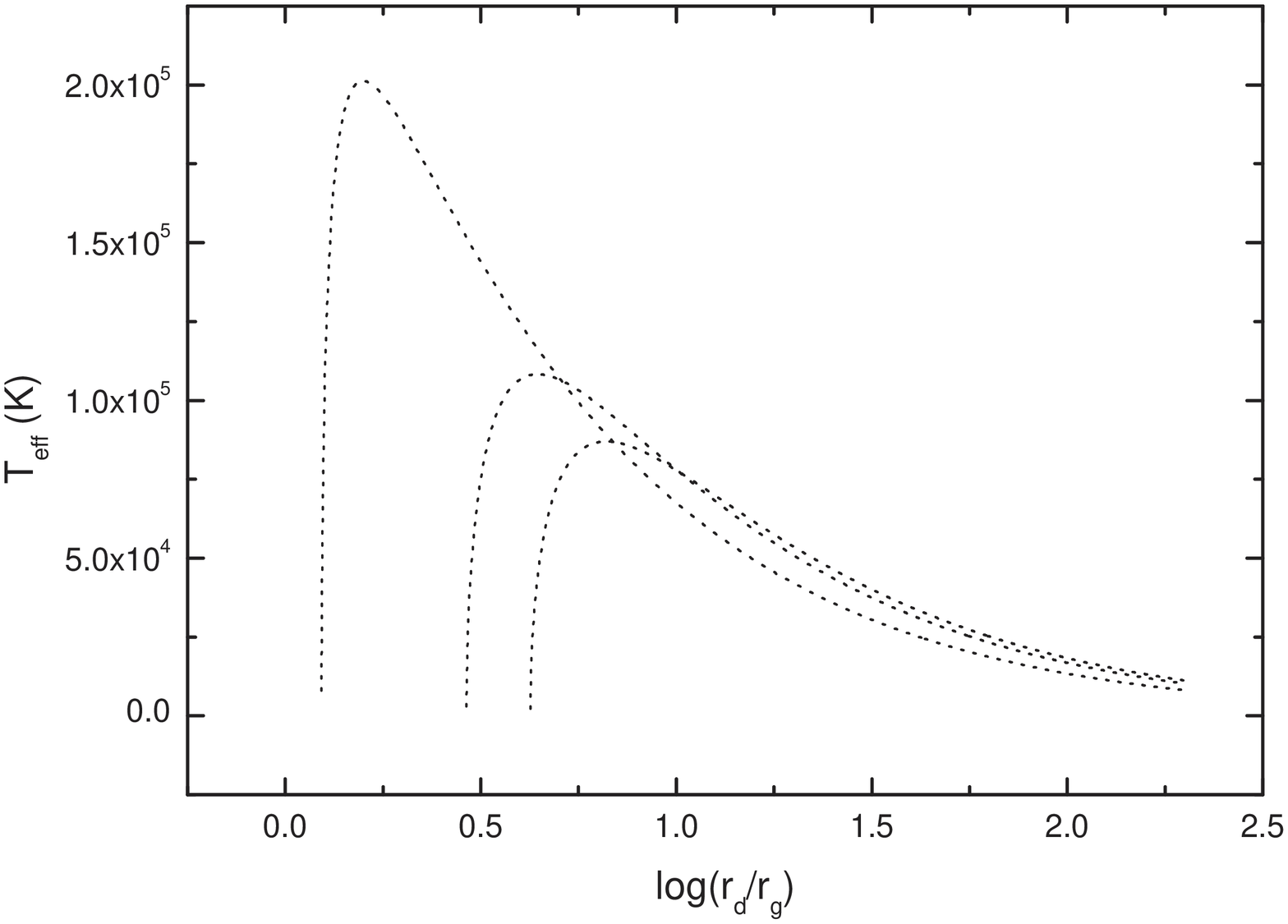}
 \caption{Radial profiles of $T_{\rm{eff}}$ shown in the basis of the standard accretion disk under
 the Kerr potential for 3C 279. The curves are calculated by adopting $f_{\rm{cov}}=0.03$, $L_{\rm{BLR}}=10^{44.41}\/\
 \rm{ergs \/\ s^{-1}}$, and $M_{\rm{BH}}=10^{8.4} \/\ M_{\sun}$. From the top down, the curves correspond
 to $a_{\rm{\ast}}=0.998$, $a_{\rm{\ast}}=0.8$, and $a_{\rm{\ast}}=0.5$, respectively.}
 \label{fig1}
\end{figure}

\begin{figure}
\centering
\includegraphics[angle=-90,scale=0.5]{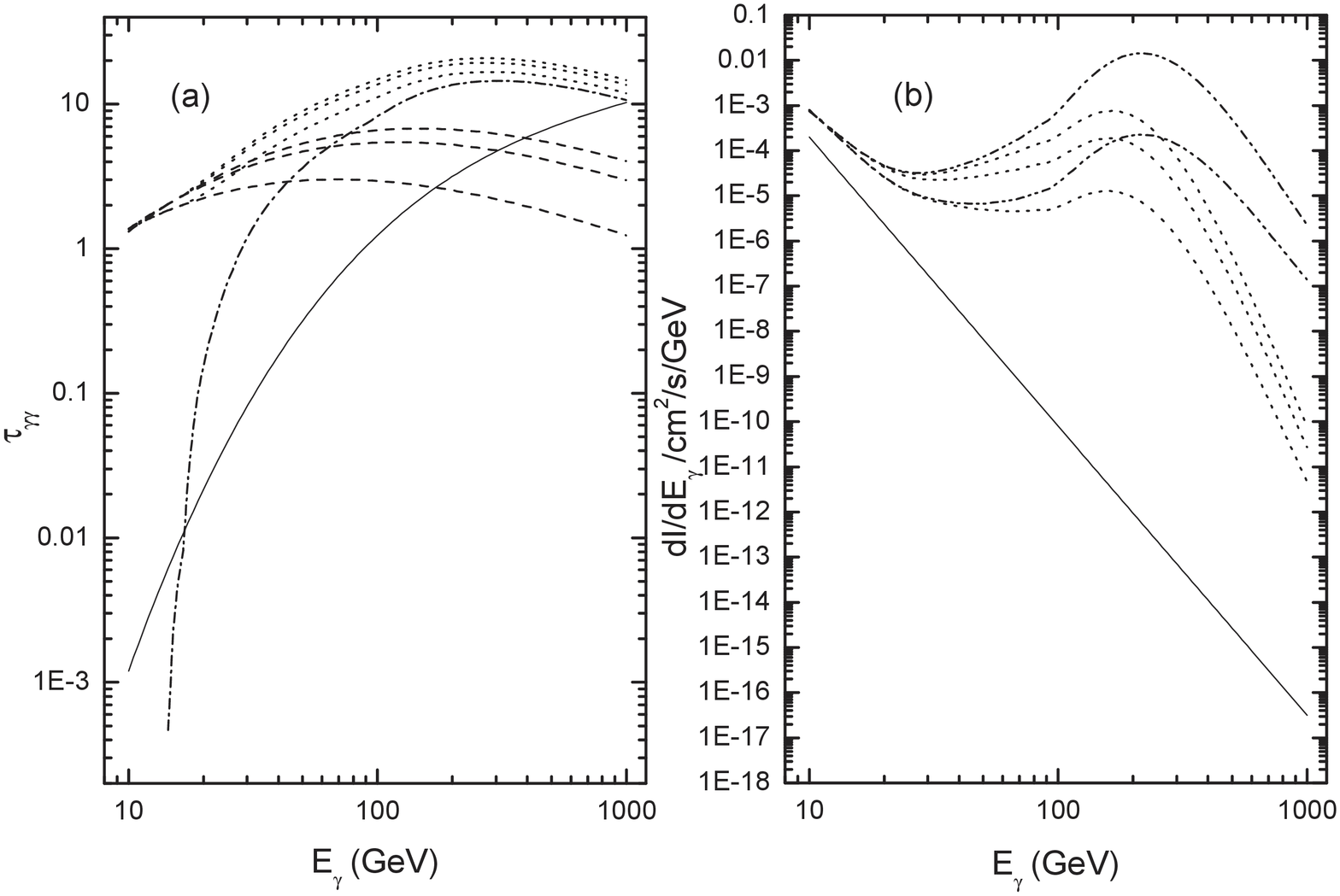}
 \caption{Plot $(a)$ is photon-photon absorption optical depth for 3C 279. Dashed lines are absorption by
 multi-temperature blackbody, dash-dotted line is absorption by broad emission lines, and dotted lines are the total
 absorption by the two radiation fields. In calculations of $\tau_{\gamma\gamma}$, we assumed
 $R_{\rm{\gamma}}=r_{\rm{BLR,in}}$, $\tau_{\rm{BLR}}/f_{\rm{cov}}=1$, and $f_{\rm{cov}}=0.03$, and adopted
 $r_{\rm{BLR,in}}=0.1 \/\ \rm{pc}$, $r_{\rm{BLR,out}}=0.4 \/\ \rm{pc}$, $L_{\rm{BLR}}=10^{44.41}\/\
 \rm{ergs \/\ s^{-1}}$, and $M_{\rm{BH}}=10^{8.4} \/\ M_{\sun}$. Plot $(b)$ is
 gamma-ray spectrum of 3C 279: the spectrum expressed by equation (1) (solid line), and the corrected spectra for
 photon-photon absorption by the diffuse radiation fields of BLR (dotted lines). From the top down, dotted and dashed
 curves in plot $(a)$ are calculated by assuming $a_{\rm{\ast}}=0.5$, $a_{\rm{\ast}}=0.8$, and $a_{\rm{\ast}}=0.998$,
  respectively. From the top down, dotted curves in plot $(b)$ are calculated by assuming $a_{\rm{\ast}}=0.998$,
  $a_{\rm{\ast}}=0.8$, and $a_{\rm{\ast}}=0.5$, respectively. Solid line in plot $(a)$ is the external absorption
  due to the EBL, estimated by equation (4). Dash-dot-dotted
  lines in plot $(b)$ correspond to the dotted lines in plot $(a)$, in the case of $a_{\rm{\ast}}=0.5$ and
  $a_{\rm{\ast}}=0.998$, but corrected for the EBL absorption.
  }
 \label{fig2}
\end{figure}

\begin{figure}
\centering
\includegraphics[angle=-90,scale=0.5]{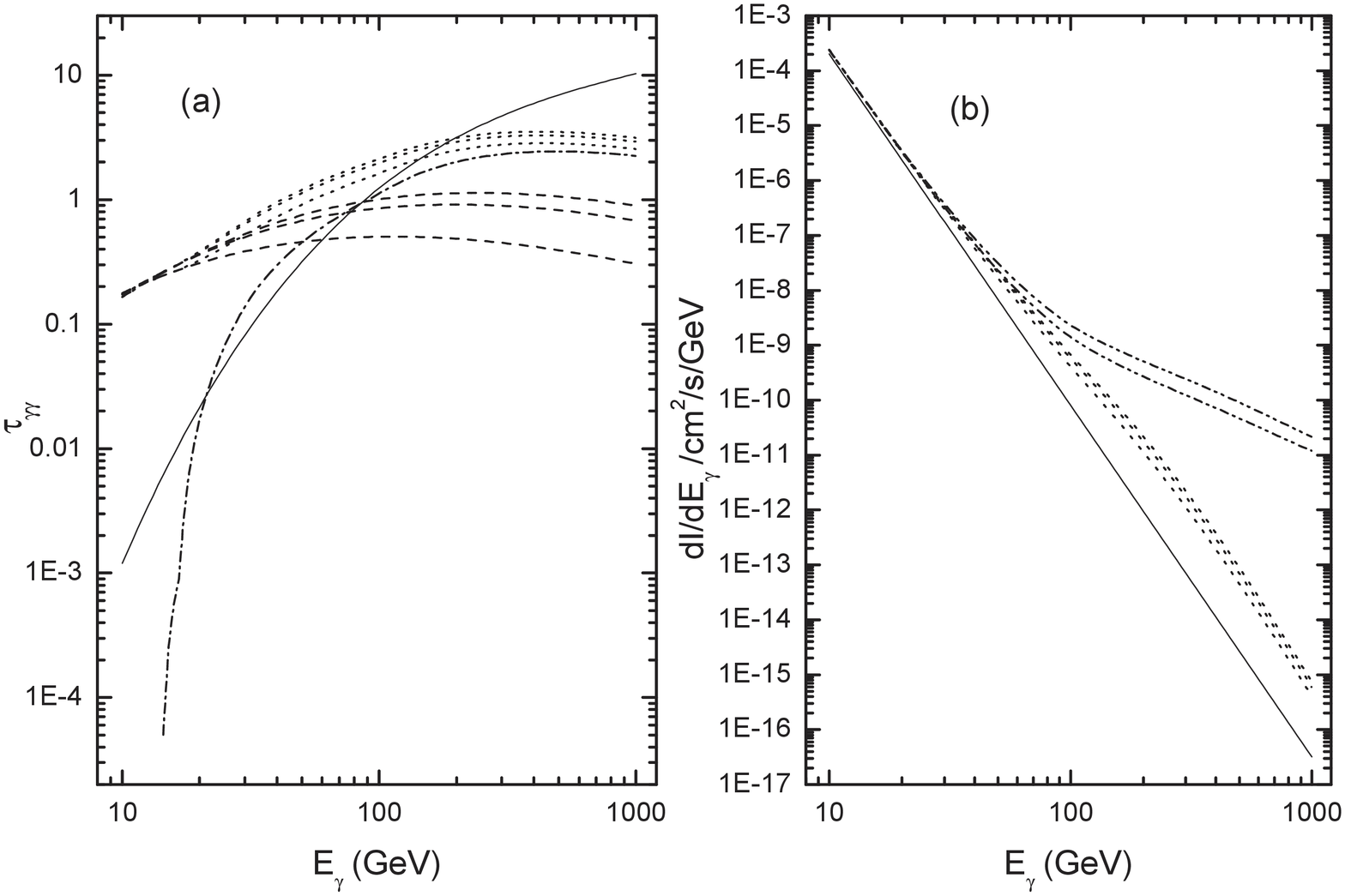}
 \caption{The patterns of lines are the same as ones in Figure 2 with the same parameters except for $R_{\rm{\gamma}}=(r_{\rm{BLR,in}}+r_{\rm{BLR,out}})/2=0.25 \/\
 \rm{pc}$.}
\label{fig3}
\end{figure}

\begin{figure}
\centering
\includegraphics[angle=-90,scale=.5]{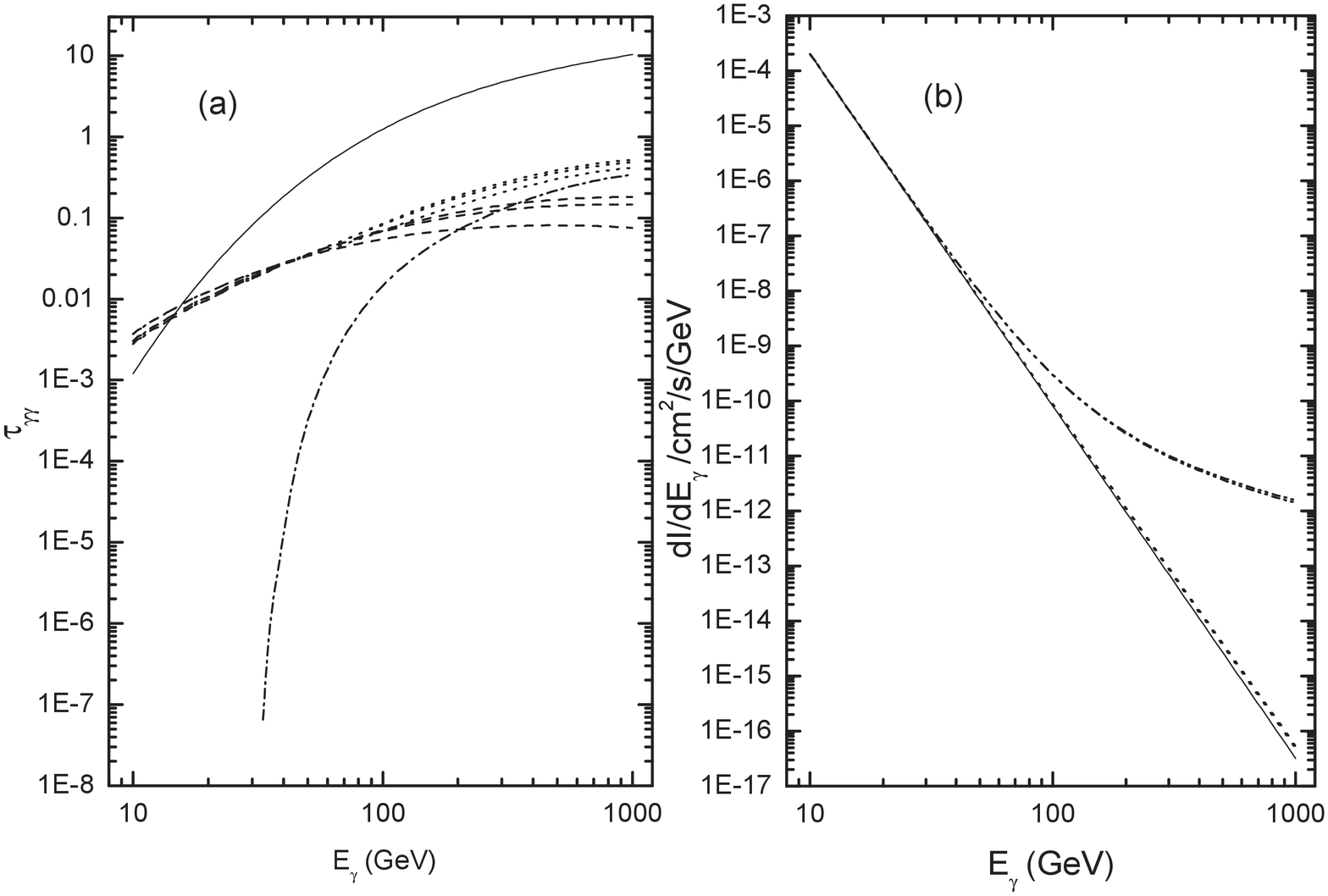}
 \caption{The patterns of lines are the same as ones in Figure 2 with the same parameters except for
 $R_{\rm{\gamma}}=r_{\rm{BLR,out}}$.}
  \label{fig4}
\end{figure}

\begin{figure}
\centering
\includegraphics[angle=-90,scale=.5]{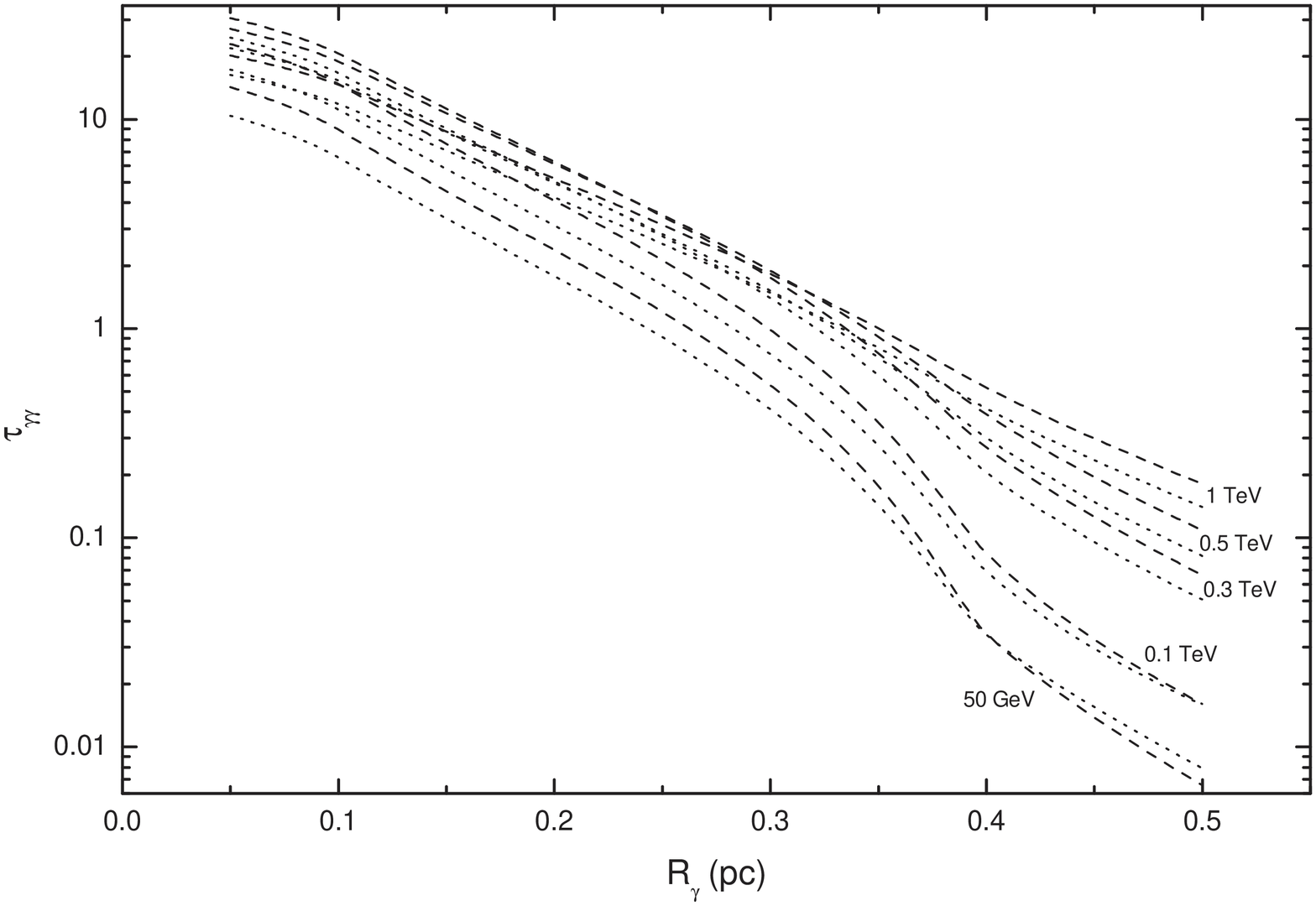}
 \caption{Dependence of $\tau_{\gamma\gamma}$ on $R_{\gamma}$. Dotted lines are the absorption optical depth
 in the case of $a_{\rm{\ast}}=0.998$, and dashed lines $a_{\rm{\ast}}=0.5$.}
  \label{fig5}
\end{figure}

\begin{figure}
\centering
\includegraphics[angle=-90,scale=.5]{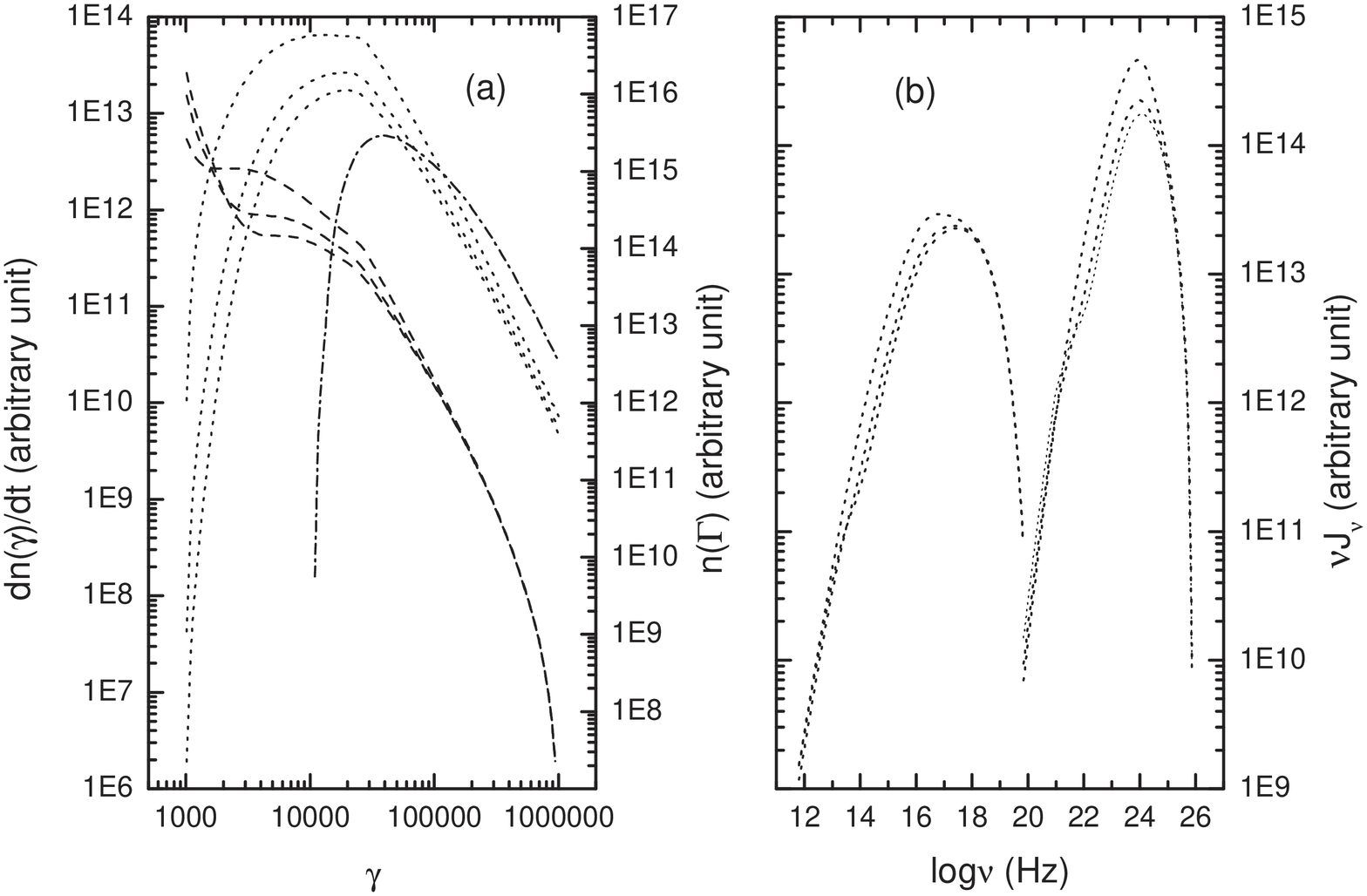}
 \caption{Pair production spectrum from photon-photon annihilation. Plots show $(a)$ the differential
 pair production rate produced by gamma rays (assumed as $\propto E_{\gamma}^{-2.3}$)
 and multi-temperature blackbody (dotted curves), the one
 produced by gamma rays and broad emission lines (dash-dotted curve), and the total equilibrium pair distribution
 for the two pair production rate (dashed curves); $(b)$ the synchrotron and EC spectra emitted by the equilibrium
 pair distribution. The curves are calculated by adopting $f_{\rm{cov}}=0.03$, $L_{\rm{BLR}}=10^{44.41}\/\
 \rm{ergs \/\ s^{-1}}$, and $M_{\rm{BH}}=10^{8.4} \/\ M_{\sun}$. From the top down, dotted and dashed
 curves correspond to $a_{\rm{\ast}}=0.998$, $a_{\rm{\ast}}=0.8$, and $a_{\rm{\ast}}=0.5$, respectively.}
  \label{fig6}
\end{figure}

\begin{figure}
\centering
\includegraphics[angle=-90,scale=.5]{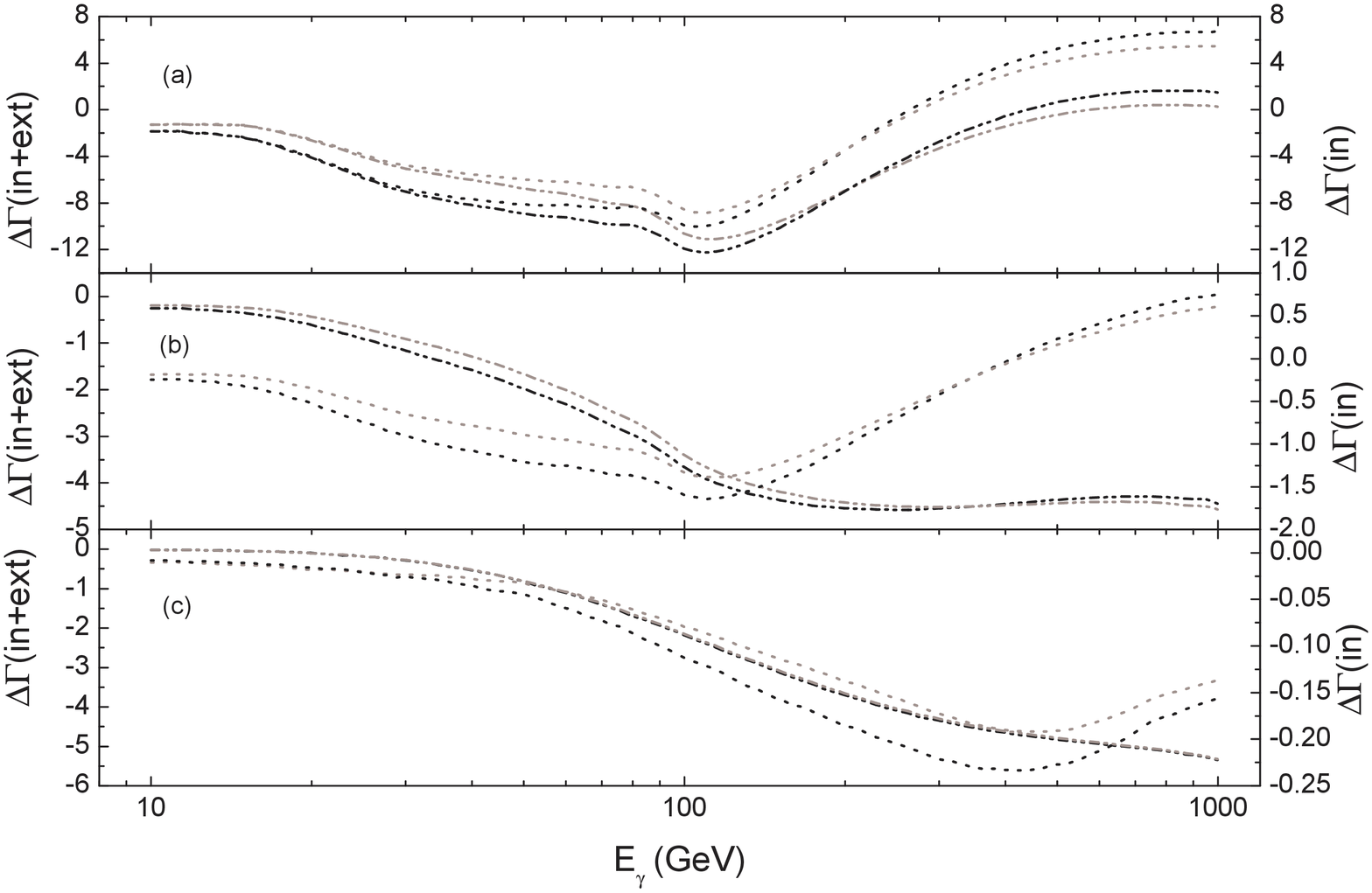}
 \caption{Photon index variations of gamma-ray spectra presented in Figures 2$b$, 3$b$, and 4$b$. $(a)$
 $R_{\gamma}=r_{\rm{BLR,in}}$. $(b)$ $R_{\gamma}=(r_{\rm{BLR,in}}+r_{\rm{BLR,out}})/2$. $(c)$ $R_{\gamma}=r_{\rm{BLR,out}}$.
Gray curves are calculated by adopting $a_{\rm{\ast}}=0.998$, and
black curves $a_{\rm{\ast}}=0.5$. Dotted curves present gamma-ray
spectra corrected for the internal absorption, and dash-dot-dotted
curves these corrected for the internal and external absorption.
 }
  \label{fig7}
\end{figure}

\end{document}